\documentclass[a4paper]{jpconf}
\usepackage{graphicx}
\usepackage{amsmath}
\begin{document}
\title{The search for Light Dark Matter with NEWS-G}

\author{Daniel Durnford and Marie-Cécile Piro on behalf of the NEWS-G Collaboration}
\address{Department of Physics, University of Alberta, Edmonton, Alberta, T6G 2R3, Canada}
\ead{ddurnfor@ualberta.ca, mariecci@ualberta.ca}

\begin{abstract}
The NEWS-G direct dark matter search experiment uses spherical proportional counters (SPC) with light noble gases to explore low WIMP masses. The first results obtained with an SPC prototype operated with Ne gas at the Laboratoire Souterrain de Modane (LSM) have already set competitive results for low-mass WIMPs. The forthcoming next phase of the experiment consists of a large 140 cm diameter SPC installed at SNOLAB with a new sensor design, with improved detector performance and data quality. Before its installation at SNOLAB, the detector was commissioned with pure methane gas at the LSM, with a temporary water shield, offering a hydrogen-rich target and reduced backgrounds. After giving an overview of the improvements of the detector, preliminary results of this campaign will be discussed, including UV laser and Ar-37 calibration data.
\end{abstract}

\section{Introduction}
\label{sec:intro}
Astronomical and cosmological observations provide strong evidence that most of the matter in the Universe is in the form of cold, non-baryonic dark matter \cite{clowe,planck}. The best-motivated candidate is a stable, massive, neutral particle generically referred to as a WIMP (Weakly Interacting Massive Particle) \cite{feng}. Direct detection experiments typically look for nuclear recoil signatures in the range of (1–100) keV coming from the elastic scattering of WIMPs off a target nucleus with masses of (10–1000) GeV/c$^2$. In the last decade, new theoretical motivation for low mass WIMPs searches (GeV/c$^2$ range) has increased interest in the community, but it requires sub-keV recoil energy thresholds and detector materials containing low mass nuclei. New Experiments With Spheres-Gas (NEWS-G) is a dark matter direct-detection experiment using spherical proportional counters \cite{og_spc} (SPCs) with light noble gases and has been designed to be specifically sensitive to detect very low WIMP masses down to 0.1 GeV/c$^2$. The first results in 2017 from the NEWS-G collaboration, obtained with a 60-cm diameter prototype SPC, SEDINE, operated in a compact radiation shielding with a Ne + CH4 (0.7\%) at 3.1 bars for a total exposure of 9.7 kg-days at the Laboratoire Souterrain de Modane (LSM) \cite{lsm} set new constraints on the spin-independent WIMP-nucleon scattering cross-section for WIMP masses under 0.6 GeV/c$^2$ \cite{newsg_first}. The 42.7 day-long physics run taken between in 2015 confirmed the relevance of SPCs for dark matter searches and placed NEWS-G as a leader in the search for low-mass WIMPs. The current phase of the experiment consists of a larger sphere of 140 cm diameter SPC to be operated at SNOLAB \cite{snolab}. However, a first campaign has been performed with pure methane gas at the LSM, a temporary water shield, and improvements in the reduction of the background levels. The use of a lighter target, improved threshold, and detector performance will allow for unprecedented sensitivity to sub-GeV WIMPs down to 0.1 GeV/c$^2$.

\section{NEWS-G Detector}
\label{sec:detector}
\begin{figure}[h]
\includegraphics[width=0.65\textwidth]{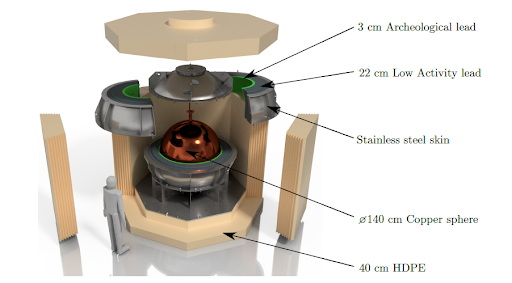}\hspace{2pc}%
\begin{minipage}[b]{11pc}\caption{\label{label}Schematic of the $140\,\mathrm{cm}\;$ diameter detector and shielding setup.}
\label{fig:detector}
\end{minipage}
\end{figure}

The NEWS-G collaboration has built a 140-cm diameter sphere made with ultra-low background copper (C10100) and radio-pure material shielding to minimize backgrounds that come from radioactive impurities within detector construction materials. The SPC is placed in a concentric 25 cm low-radioactivity lead shield, with the innermost 3 cm of lead made from ultra-low $^{210}$Pb content archaeological lead and surrounded by 40 cm of high-density polyethylene to shield against neutrons from the environment. The inter-space is continuously flushed with pure nitrogen to mitigate the presence of radon. The detector schematic is shown in Figure \ref{fig:detector}.

The two C10100 hemispheres were fabricated in France. An electroplating method developed at PNNL (USA) was implemented to add 0.5 mm of pure copper to the inside of the hemispheres, to limit the impact of the $^{210}$Pb contamination found in the C10100 copper used to fabricate the SPC \cite{newsg_copper}. This work was performed underground at LSM to limit the production of cosmogenic radioactive backgrounds. The hemispheres were then welded into a sphere using an electron beam in France. After cleaning and etching the sphere underground at LSM, the sphere was kept isolated from the naturally present radon in a sealed, nitrogen-filled bag. Additional efforts have been made to improve the gas handling system and gas quality, including a series of filters to remove electro-negative impurities such as water and oxygen and radon. A radon trap system \cite{radon_patrick} has been developed at the University of Alberta, and adsorbent materials and their optimal temperature conditions are explored for the forthcoming phase at SNOLAB. In addition, a novel sensor called ACHINOS with eleven individual anodes has been developed by NEWS-G collaborators to improve charge collection in large detectors by increasing the magnitude of the electric field at large radii \cite{newsg_achinos}. The sensor was read out in two channels, the lower (``southern'') $6$ anodes and upper (``northern'') $5$ anodes. The complete installation of the 140 cm SPC with the lead shielding was achieved at LSM with a temporary neutron shield to allow the full commissioning of the detector using pure methane in low background conditions before being dismounted and shipped to SNOLAB. As with previous NEWS-G experiments, the current induced on the anode by the movement of avalanche ions is integrated and amplified with a CREMAT charge sensitive pre-amplifier \cite{cremat}, then digitized \cite{newsg_first}.

\section{Commissioning at the LSM}
\label{sec:lsm}

A limited physics campaign at the LSM lasting approximately $6$ days was carried out with the $140\,\mathrm{cm}$ SPC. Pure methane gas was used as a target material ($135\,\mathrm{mbar}$), which offers the advantage of a larger fraction of hydrogen target material, with lower backgrounds from Compton scattering. Only data collected from the southern channel of the ACHINOS sensor is used to produce physics results, as the electric field and detector response are uniform in this region of the sphere, which spans roughly $70\%$ of the volume of the detector.

\begin{figure}[h]
\includegraphics[width=0.52\textwidth]{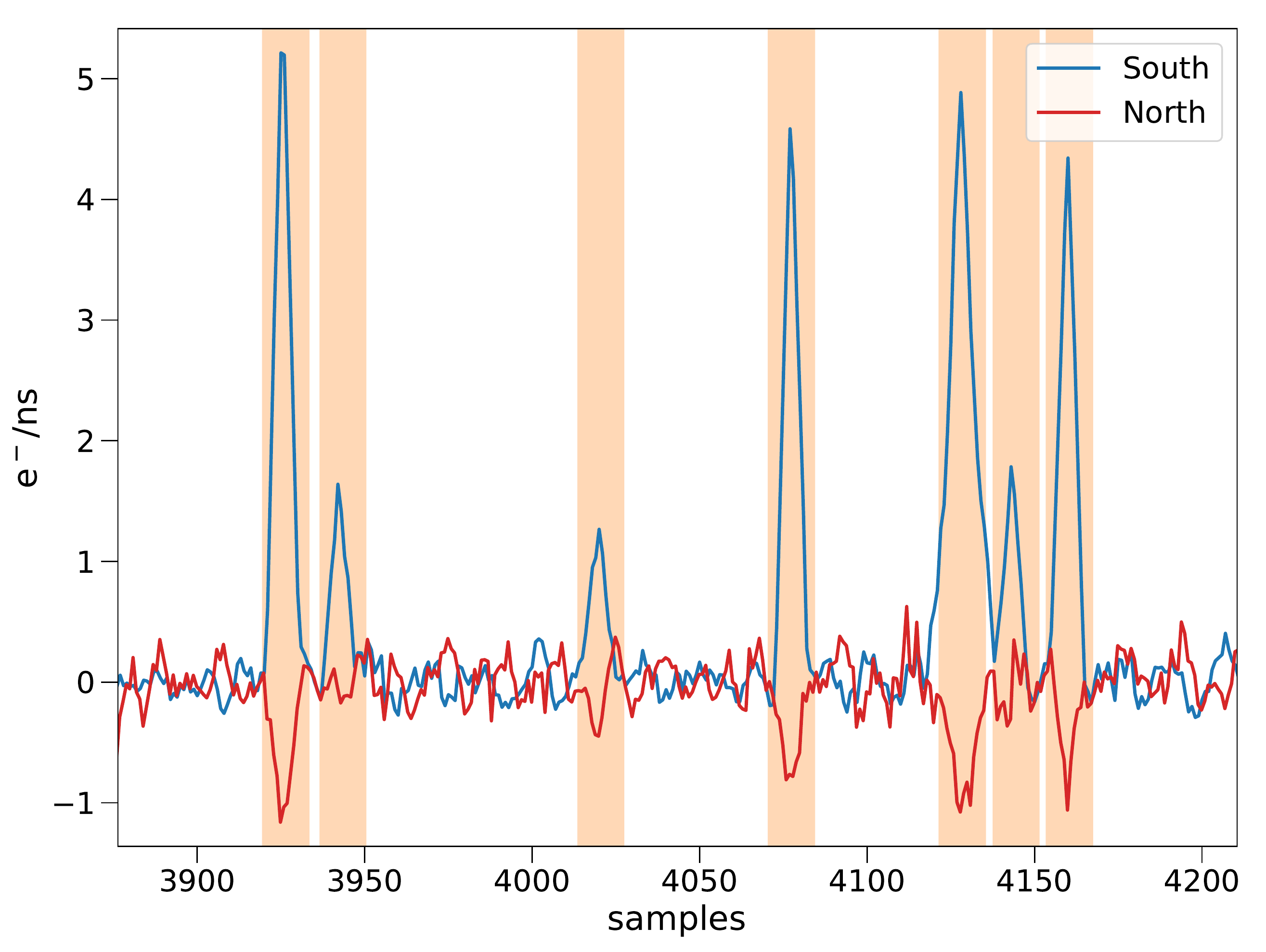}\hspace{2pc}%
\begin{minipage}[b]{14pc}\caption{\label{fig:NS}Example of a treated physical event trace, double-deconvolved to remove distortion due to the avalanche process and pre-amplifier response. The Townsend avalanche occurred at one or more anodes on the ``southern'' half of the ACHINOS sensor (each peak representing an individual avalanche event), with negative cross-talk induced on the ``northern'' ACHINOS channel.\\}
\end{minipage}
\end{figure}

Unlike previous NEWS-G experiments, the large drift volume and high-diffusion gas condition yields improved temporal resolution. Individual primary electrons can be identified in event traces deconvolved for the pre-amplifier and ion-induced current responses \cite{newsg_paco}; an example of this is shown in Fig.\ \ref{fig:NS}. To take advantage of this additional information, a peak-finding algorithm based on ROOT TSpectrum class is applied to the data \cite{root}. This allows for the rejection of background signals, many of which are single-peak events. A large portion of low energy background events observed are after-pulse signals induced by alpha decays (primarily $\mathrm{^{210}Po}$ on the surface of the sphere), which produce significant increases in the rate of single-peak events, as well as a distortion in the electric field due to the $\sim 10^8$ charges produced in the alpha particle's Townsend avalanche, observable as a decrease in electron drift time. This background is mitigated by removing $5\,\mathrm{s}$ after observed alpha decays, leading to a roughly $13\%$ loss of exposure. Another phenomenon unique to this experimental setup was the ``cross-talk'' observed between the two channels of the ACHINOS sensor. Physical pulses producing a current on one channel were found to induce a smaller, inverted signal on the other with a consistent fractional amplitude of approximately $20\%$ (see Fig.\ \ref{fig:NS}). However, non-physical pulses (that do not produce an avalanche in the gas) do not follow this trend, which is exploited as a means of discrimination against non-physical background events.

\begin{figure}[h]
\begin{minipage}{18pc}
\includegraphics[width=\textwidth]{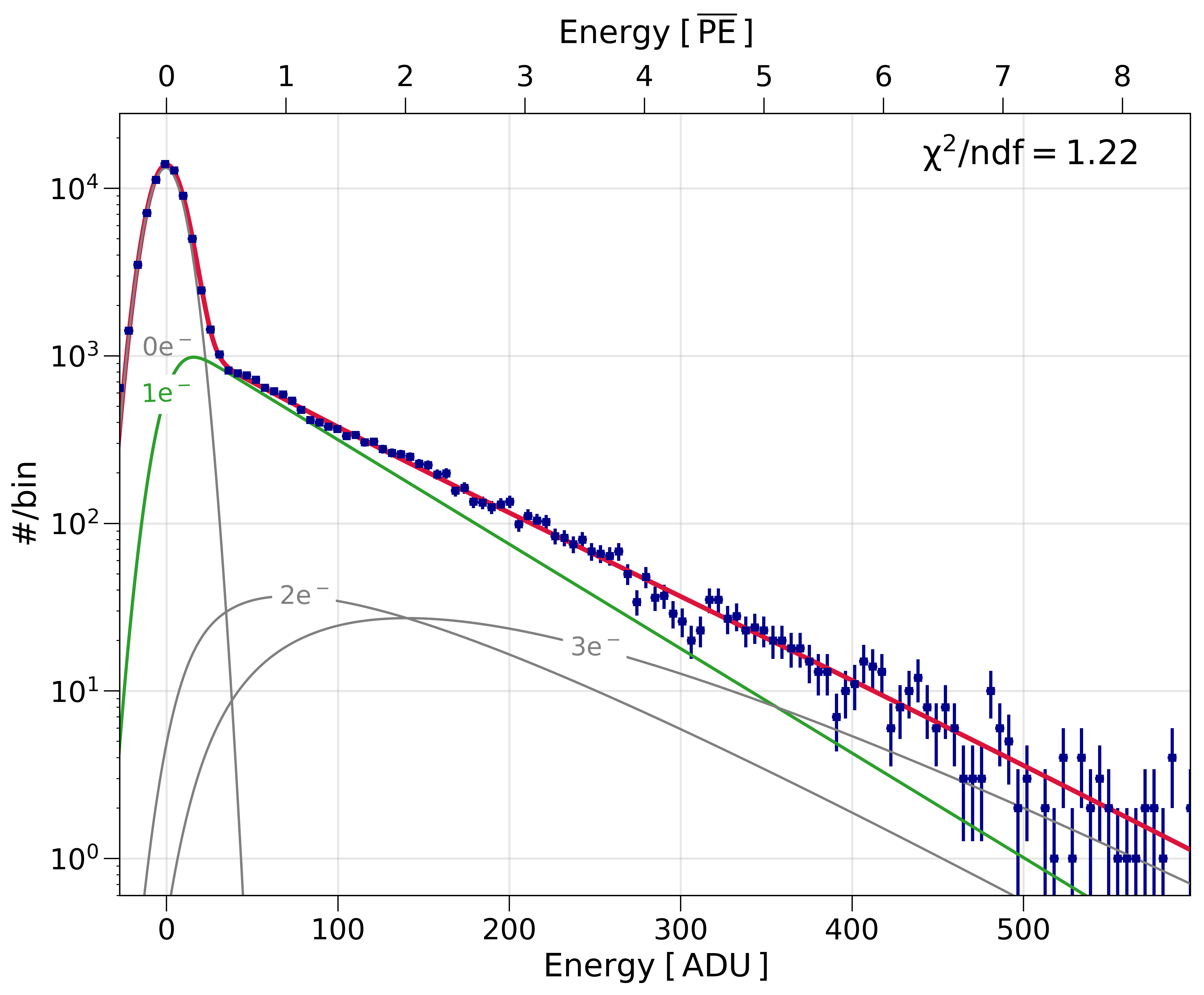}
\caption{\label{fig:laser}Fit of a laser calibration data set (blue). Components of the model for different numbers of primary electrons are labeled, with the overall fit in red. The energy scale is converted to the average number of primary electrons for the top axis scale using the best-fit value of the gain for this data.}
\end{minipage}\hspace{2pc}%
\begin{minipage}{18pc}
\includegraphics[width=\textwidth]{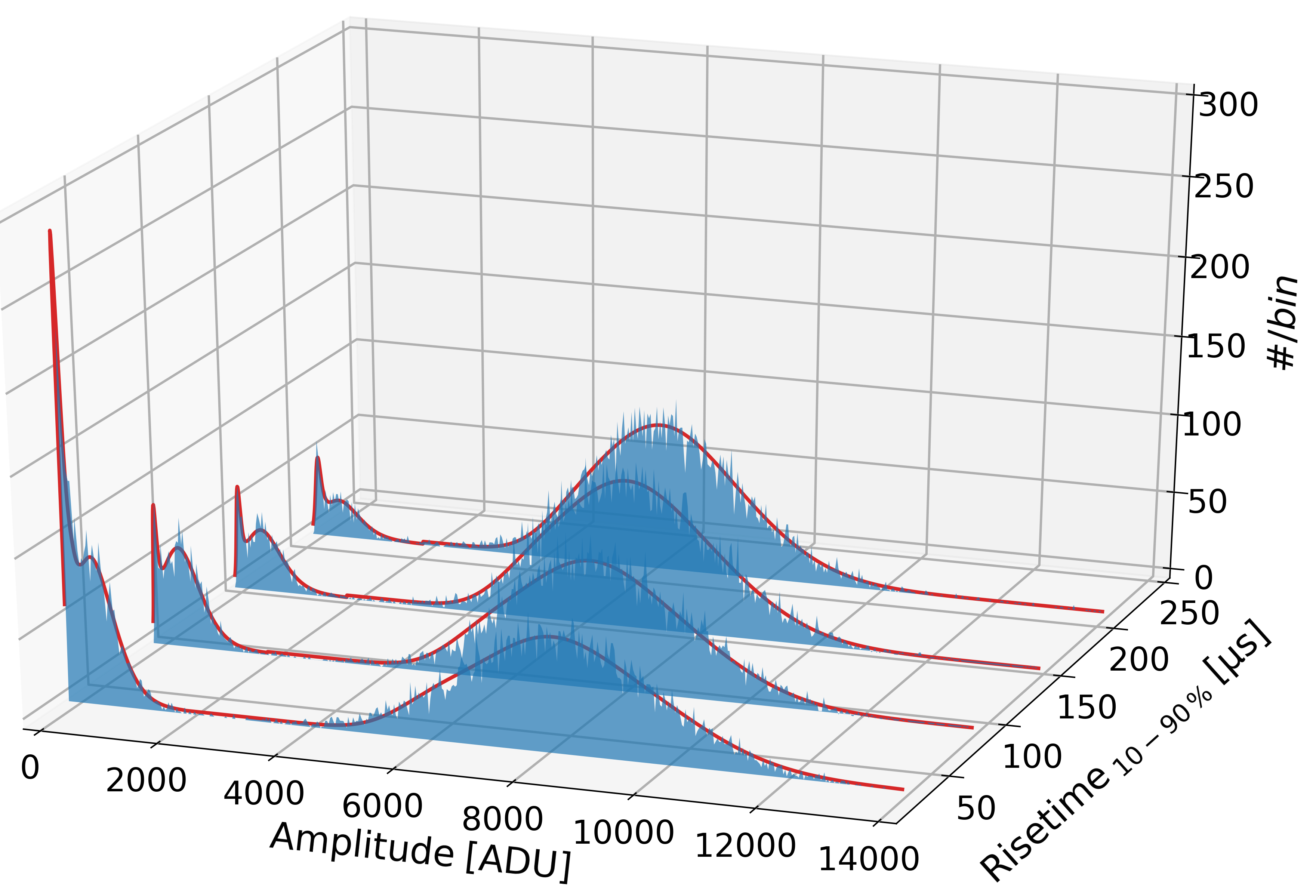}
\caption{\label{fig:ar37}Fit of the Ar-37 calibration data (blue histogram) as a function of amplitude, for several slices of risetime. This fit includes a background component (the lowest amplitude peak visible) modeled using data taken without the source present.}
\end{minipage} 
\end{figure}

During this campaign, calibration data was collected with a $213\,\mathrm{nm}$ pulsed UV laser system \cite{newsg_laser}, using a photo-diode to tag these events, in which primary electrons are extracted from the surface of the sphere via the photoelectric effect. The laser was continuously run at a pulse rate of $10\,\mathrm{Hz}$ to monitor changes in detector operating conditions, including a roughly $10\%$ drop in detector gain over the course of the campaign due to increasing gas contamination. Between daily physics runs, dedicated laser calibrations with low-intensity laser pulses occurred, producing on average approximately $5$ primary electrons. These electrons then drift to the southern hemisphere of the ACHINOS sensor, providing tagged, low-energy calibration events to characterize the SPCs avalanche response. The resulting amplitude spectrum, shown in Fig.\ \ref{fig:laser} for one calibration run, allows for characterization of the SPCs single electron response. This is modeled with the Polya distribution \cite{polya1,polya2} (a modified exponential) with shape parameter $\theta$ ($\theta=0$ is an exponential distribution, but approaches a Gaussian for large $\theta$) \cite{newsg_laser}. Additionally, the mean gain $\left \langle G \right \rangle$ of the detector can be extracted from this calibration. The peak finding algorithm is also characterized with the laser calibration data by including contributions of false negative, false positive, and coincident peaks in the amplitude spectrum of identified single electron events. The peak-finding algorithm identifies $64\%$ of primary electrons, with a $0.1\%$ chance of falsely identifying noise as peaks in a span of $60\,\mathrm{\mu s}$. Comparing the amplitude spectrum of photo-diode tagged events to those which triggered the sphere DAQ yields a hardware trigger efficiency for primary electrons of $54\%$. 

Another component of the calibration program is the use of $\mathrm{^{37}Ar}$, a radioactive gas producing two low-energy electron capture decays of $270\,\mathrm{eV}$ and $2.8\,\mathrm{keV}$ \cite{newsg_ar37}. This allows for confirmation of the linearity of the detector energy response over this energy regime, and modelling of the detector response in energy and risetime (a PSD variable proportional to radii). To fit this data, the detector's energy response was modeled using the COM-Poisson distribution for primary ionization \cite{newsg_fano}, and the Polya distribution for avalanche ionization \cite{newsg_laser}. Electro-negative contaminants in the gas (such as $\mathrm{O_2}$ and $\mathrm{H_2O}$) preferentially reduce the amplitude of events originating farther from the sensor (at higher risetimes), which is modeled using binomial statistics for the number of un-trapped primary electrons, with the survival probability varying linearly with risetime. The relative rates of $2.8\,\mathrm{keV}$ and $270\,\mathrm{eV}$ decays were determined using a Geant4 simulation \cite{bham,geant4}. Additionally, some $2.8\,\mathrm{keV}$ decays produce $2.6\,\mathrm{keV}$ x-rays which can escape the sphere, leaving $200\,\mathrm{eV}$ to be deposited by electrons. The laser calibrations provide a constraint on the gain of the southern-most anode (where laser events are collected), and the gain of the other $5$ southern hemisphere anodes was allowed to vary. The resulting fit is shown in Fig.\ \ref{fig:ar37}. This yields estimates for many detector-response parameters, including the W-value and Fano factor at several different interaction energies, and the gain of the entire southern channel of the ACHINOS sensor.

\section{Future Steps at SNOLAB and Conclusions}
\label{sec:snolab}

\begin{figure}[h]
\includegraphics[width=0.55\textwidth]{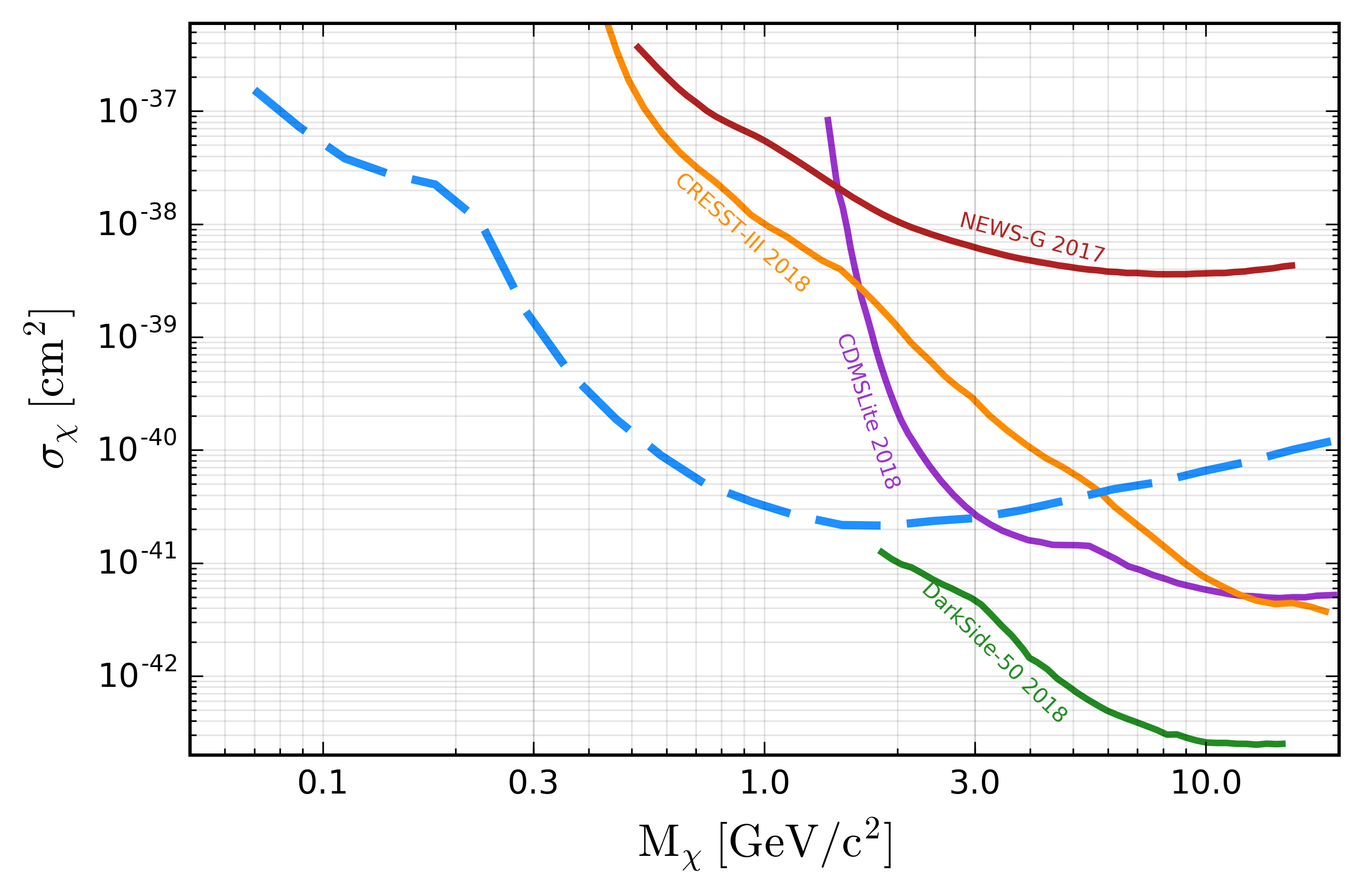}\hspace{2pc}%
\begin{minipage}[b]{14pc}\caption{\label{fig:projection}Preliminary expected sensitivity of NEWS-G at SNOLAB to spin-independent interactions (dashed blue line). Other published results are shown for comparison \cite{newsg_first,darkside,cresst,cdms}. This limit is produced with the optimum interval method \cite{yellin} assuming $20\,\mathrm{kg \cdot days}$ of exposure with $\mathrm{Ne}\,+\,10\,\% \;\mathrm{CH_4}$ gas, a flat background of $1.67\,\pm\,0.5\,\mathrm{d.r.u.}$, and a quenching factor calculated using SRIM \cite{srim}.}
\end{minipage}
\end{figure}

The $140\,\mathrm{cm}$ SPC has since been relocated to SNOLAB, with installation complete as of September 2021. After initial commissioning, physics data will be taken with a neon-methane gas mixture, yielding sensitivity down to WIMP masses of $0.1\,\mathrm{GeV/c^2}$. As was the case at the LSM, UV laser and $\mathrm{^{37}Ar}$ calibration will be used to characterize the detector's energy response at the single electron level. The projected exclusion curve for spin-independent interactions for this experiment is shown in Fig.\ \ref{fig:projection}. In summary, NEWS-G has successfully operated a new $140\,\mathrm{cm}$ SPC at the LSM with methane gas. Dark matter results are expected from this campaign soon. With the demonstrated capability to observe single ionization quanta, NEWS-G is expected to prove highly sensitive to nuclear recoils from low mass dark matter, both with existing data in methane gas and upcoming work at SNOLAB.

\section*{Acknowledgements}

This research was undertaken, in part, thanks to funding from the Canada Excellence Research Chairs Program, the Canada Foundation for Innovation, the Arthur B. McDonald Canadian Astroparticle Physics Research Institute, Canada, the French National Research Agency (ANR-15-CE31-0008), and the Natural Sciences and Engineering Research Council of Canada.


\end{document}